\documentclass[superscriptaddress,nofootinbib,twocolumn]{revtex4}
\usepackage{amsmath,lscape,epsfig}
\usepackage{amsfonts}
\usepackage{amsmath}
\usepackage{amssymb}
\usepackage{slashed}
\usepackage{color,soul}
\usepackage[utf8]{inputenc}
\usepackage{graphicx}
\usepackage{epstopdf}
\usepackage{ulem}

\def\beq{\begin{equation}}
\def\eeq{\end{equation}}
\def\beqa{\begin{eqnarray}}
\def\eeqa{\end{eqnarray}}
\def\ban{\begin{eqnarray*}}
\def\ean{\end{eqnarray*}}
\def\bi{\begin{itemize}}
\def\ei{\end{itemize}}

\begin{document}

\title{Schwinger mechanism in the SU(3) Nambu–Jona-Lasinio model
with an electric field}

\author{William R. Tavares} \email{william.tavares@posgrad.ufsc.br}
\affiliation{Departamento de F\'{\i}sica, Universidade Federal de Santa
  Catarina, 88040-900 Florian\'{o}polis, Santa Catarina, Brazil}  

\author{Sidney S. Avancini} \email{sidney.avancini@ufsc.br}
\affiliation{Departamento de F\'{\i}sica, Universidade Federal de Santa
  Catarina, 88040-900 Florian\'{o}polis, Santa Catarina, Brazil}

\begin{abstract}

In this work we study the electrized quark matter under finite temperature 
 and density conditions in the context of the SU(2) and SU(3) Nambu-Jona-Lasinio models.
To this end, we evaluate the effective quark masses and 
the Schwinger quark-antiquark pair production rate. 
For the SU(3) NJL model 
we incorporate in the Lagrangian the 't Hooft determinant and we present a set of 
analytical expressions more convenient for numerical evaluations. We predict a decrease of the pseudocritical
electric field with the increase of the temperature for both models and a more prominent production 
rate for the SU(3) model when compared to the SU(2). 
 \vspace{0.5cm}
 \\

\end{abstract}

\maketitle

\vspace{0.50cm}
PACS number(s): {12.39.-x, 12.38.-t, 13.40.-f, 12.20.Ds}
\vspace{0.50cm}

\section{Introduction}

In the last few decades strongly interacting quark matter under extreme conditions of 
temperature and/or baryon density has been 
extensively studied due not only to the possibility of a phase transition from hadronic matter 
to the quark-gluon-plasma(QGP), 
but also the possibility for exploiting properties of the fundamental interactions. Such conditions 
are explored in
accelerators like LHC-CERN and BNL-RHIC, and also can be found in compact objects like neutron 
stars\cite{neutron} or in the
early universe\cite{yagi,vachaspati}.

To study this type of matter under such conditions in the low energy sector of Quantum
Chromodynamics(QCD) becomes hard to handle 
and Lattice QCD simulations are limited
due to the sign problem \cite{Muroya}. One of the most common approaches is to use effective theories.
In this scenario, a phase diagram of the transition from the hadronic matter to QGP can be plotted, 
and it is expected that exists a crossover at high temperatures and low baryonic densities;
otherwise, a first-order phase transition at high densities and low temperatures. At even higher 
baryonic densities it is expected
a color superconducting phase\cite{fukushima}. A natural extension is the introduction of strong magnetic 
fields, that has been calling the attention 
due to the possibility of generating such fields in the non-central
ultrarelativistic Heavy Ion Collisions\cite{fukushima2} with fields of the order $eB\sim10^{19}$G and 
also in some types of 
neutron stars like magnetars with surface magnetic fields of the order $eB\sim10^{15}$G\cite{duncan,kouve}.

The chiral condensate guides the chiral symmetry restoration as an order parameter of QCD 
matter\cite{fukushima}.
Most of the effective models predictions at $B\neq0$ indicates an enhancement of chiral condensate even at $T\neq0$ and this is the phenomenon 
called magnetic catalysis\cite{shovkovy,nosso1,avancini2}. 
However, recent Lattice QCD simulations 
show a suppression of the condensates at magnetic fields of 
the order $eB \geqslant  0.2$GeV$^2$ at $T\sim T_c$\cite{bali}, i.e., a phenomenon named inverse magnetic 
catalysis which is 
not fully understood and not predicted in most of the effective theories.

Simulations using event-by-event fluctuations of the proton positions in the colliding nuclei 
in Au$+$Au heavy-ion collisions at $\sqrt{s}=200$GeV 
and in Pb$+$Pb at $\sqrt{s}=2.76$TeV,  both scales at RHIC and LHC energies, 
indicates that not only the magnetic fields already
mentioned are created, but also strong electric fields of
the same order of magnitude\cite{skokov,deng,zhang,zhang2}. 
 Besides, in asymmetric Cu$+$Au collisions\cite{hirono,deng2,Voronyuk} it is predicted that a strong electric
field is generated in the overlapping region.
\cite{hirono,deng2,Voronyuk}. This happens because there is a 
different number of electric charges in each nuclei, 
and it is argued that this is a fundamental property
due to the charge dipole formed in the early stage of the collision.
 Recently, extensive efforts have been done to study the 
chiral magnetic effect\cite{fukushima}.  However, it is expected in the case where external electric fields 
are present, 
the Chiral Electric Separation\cite{xu1,xu2}  effect to take place, in this way
probing anomalous transport properties of the matter generated in the QGP dynamics. 

 Only few works are dedicated to explore the effects of electric fields in the chiral phase transition
\cite{Cao,ruggieri1,ruggieri2,kleva,tiburzi,yamamoto, babansky, cohen,huang2,tatsumi} in the strongly 
interacting quark matter. At $T=\mu=0$, the 
effect of pure electric fields is to restore the chiral symmetry,
although in this case we are dealing with a unstable
vacuum and with the possibility of creating quark-antiquark 
pairs of particles through
the Schwinger mechanism\cite{Schwinger,heisenberg}. As mentioned in \cite{Cao}, the estimated number of
charged quark-antiquark pairs produced in the heavy-ion
collisions with Au$+$Au and with Pb$+$Pb is quite significant, indicating that the creation of the pair of
particles should be relevant.

Our objective in this work is to consider temperature, chemical potential
and electric field in the 
context of the SU(3) and SU(2) Nambu--Jona-Lasinio Model
\cite{nambu,buballa} and study 
how the constituent quark masses and the Schwinger pair-production\cite{Schwinger,heisenberg} are altered
under the change of such variables.
\\
 Our main contributions in this paper are to update and to extend previous works devoted to the study of
electrized quark systems, but now including the 't Hooft interaction and describing in a more systematic 
way the strange quark sector. Here, we emphasize the importance of a proper regularization scheme, which
has been overlooked in some works. The experience gained from magnetic systems, which are closely related 
to the electric ones, shows that it is of fundamental importance the choice of the regularization for
obtaining results that make sense. 
We use the analytic continuation technique in order to obtain  analytical expressions for the 
effective potential and gap equation in strongly electrized systems starting from the corresponding 
regularized magnetic expressions. Now, to the best of our knowledge, these results are not given in
the literature in the present context. Often, in the literature the real part  contribution for the
gap and effective potential are obtained through the numerical calculation of the principal value of
the corresponding divergent expressions, which is cumbersome from the numerical point of view.  
Our analytical expressions circumvents these problems and give expressions very simple and easy
to be used in numerical calculations. 
In the  section II we start by presenting the formalism of the SU(3) NJL model and the principal
equations whose details will be left to the appendix. In the section III we present the 
regularization adopted in this work. Section IV 
we develop the SU(2) NJL model. In the Section V
we present our  numerical results. Finally, in section VI the conclusions are discussed.

\section{General formalism}
We start by considering the general three-flavor NJL model Lagrangian in the 
presence of a electromagnetic field 

\begin{multline}
\mathcal{L}=\overline{\psi}\left(i \slashed D - \tilde{m}\right)\psi -\frac{1}{4}F^{\mu\nu}F_{\mu\nu} \\
+G\sum_{a=0}^8\left[(\overline{\psi}_f\lambda^a\psi_f)^{2}+
(\overline{\psi}_fi\gamma_{5}\lambda^a\psi_f)^{2}\right] \\
-K \{ \det \left[ \overline{\psi}_f(1+\gamma_5)\psi_f \right] + 
\det \left[ \overline{\psi}_f(1-\gamma_5)\psi_f\right]  \} ~,\label{su3}
\end{multline}

\noindent where $A^\mu$, $F^{\mu\nu} = \partial^\mu A^\nu - \partial^\nu A^\mu$ 
are respectively the electromagnetic gauge field potential and  field tensor, 
$G$ and $K$ are the coupling 
constants, $\lambda^a$ with $a=1,...8$ are the Gell-Mann matrices and $\lambda^0=\sqrt{2/3}\mathbb{I}$, $Q$ is the diagonal quark charge, $Q=$diag$(q_u=2e/3,q_d=-e/3,q_s=-e/3)$,
and
$D^\mu =(i\partial^{\mu}-QA^{\mu})$ is the covariant derivative.   
 The quark fermion field is represented by $\psi_f=(u,d,s)^T$ with $f$ indicating their respective flavors and $\tilde{m}=$diag$_f(m_u,m_d,m_s)$ is the corresponding 
 (current)quark mass matrix.
 We choose $A_{\mu}=-\delta_{\mu 0}x_{3}E$ 
  to obtain a resulting constant electric field in the z-direction.

 The Lagrangian (\ref{su3}) contains scalar and pseudo-scalar four-point interactions 
and the 't Hooft determinant six-point interaction, added 
to break the U(1) symmetry\cite{buballa}. From here on, we adopt the mean field approximation, where a set of self-consistent gap equations 
are obtained through the linearization of the four and six-point interactions in eq.(\ref{su3}), yielding the 
result for the effective quark masses $M_i$ \cite{hatsuda}
\begin{equation}
 M_i=m_i-4G\phi_i+2K\phi_j\phi_k, \label{gapsu3}
\end{equation}
\noindent where in the last equation $(i,j,k)$ stands for any permutation among the flavors $(u,d,s)$.

More details can be found in refs.\cite{avancini2,hatsuda}.

Also, we will use the following definition for the condensate for each flavor $f$:

\begin{eqnarray}
\phi_f=\left \langle \overline{\psi}_f\psi_f \right \rangle =-\int\frac{d^4p}{(2\pi)^4} 
Tr\left[iS_f(p)\right].\label{cond1}
\end{eqnarray}
Since we are working in an electrized medium at finite temperatures and densities, we can subdivide $\phi_f$ 
 into a contribution with a pure electric field $\phi_f^{\mathcal{E}}$ and a thermo-eletric part $\phi_f^{\mathcal{E},T,\mu}$ 
\begin{equation}
 \phi_f=\phi_f^{\mathcal{E}}+\phi_f^{\mathcal{E},T,\mu}.
\end{equation}

Following the steps
 of \cite{Cao}, one can use the full quark propagator in a constant electric field 
using the
Schwinger proper-time method\cite{Schwinger} in order to calculate the condensate $\phi_f$ 
which reads

\begin{equation}
 \phi_f^{\mathcal{E}}=-\frac{M_fN_c}{4\pi^2}\mathcal{E}_f\int_0^{\infty}ds\frac{e^{-sM^2_f}}{s}
 \left[\cot(\mathcal{E}_f s)\right] ~, \label{phisu3}
\end{equation}
where $\mathcal{E}_f=|q_f|E$.

The thermo-electric contribution can be calculated using the Third Elliptic Theta Function\cite{Cao,ruggieri1}, and is given by
\begin{eqnarray}
 &&\phi_f^{\mathcal{E},T,\mu}=-\frac{M_fN_c}{2\pi^2}\sum_{n=1}^{\infty}(-1)^n\mathcal{E}_f\int_0^{\infty}ds\frac{e^{-sM^2_f}}{s}\cot(\mathcal{E}_fs)\nonumber\\
 &&\times e^{-\frac{\mathcal{E}_fn^2}{4|\tan(\mathcal{E}_fs)|T^2}}\cosh\left(\frac{n\mu}{T}\right),\label{phi_tmu}
\end{eqnarray}

The Thermodynamical or effective potential is necessary to evaluate the Schwinger pair-production. In the mean field approximation reads\cite{avancini2} 

\begin{equation}
 \Omega=-\theta_u-\theta_d-\theta_s+2G(\phi^2_u+\phi^2_d+\phi^2_s)-4K\phi_u\phi_d\phi_s,\label{omega1}
\end{equation}

\noindent where an irrelevant constant was discarded. We  split $\theta_f$ as
$\theta_f=\theta_f^\mathcal{E}+\theta_f^{\mathcal{E},T,\mu}$ in the same  way  as
we did with $\phi_f$, with a contribution of pure electric field and a thermo-electric part, and it is easy to show that one obtains: 
\begin{eqnarray}
 &&\theta_f^\mathcal{E}=-\frac{N_c}{8\pi^2}\int_0^{\infty}ds\frac{e^{-sM_f^2}}{s^2}\mathcal{E}_f\cot(\mathcal{E}_fs)\label{thetaE}\\
 &&\theta_f^{\mathcal{E},T,\mu}=-\frac{N_c}{4\pi^2}\sum_{n=1}^\infty(-1)^n\int_0^{\infty}ds\frac{e^{-sM_f^2}}{s^2}\mathcal{E}_f\cot(\mathcal{E}_fs)\nonumber\\
 &&\times e^{-\frac{\mathcal{E}_fn^2}{4|\tan(\mathcal{E}_fs)|T^2}}\cosh\left(\frac{n\mu}{T}\right)\label{thetaET}
\end{eqnarray}

The Schwinger pair production rate is given by $\Gamma=-2\Im\left(\Omega\right)$\cite{Cao,Schwinger}, where 
$\Im\left(\Omega\right)$ corresponds to the imaginary part of the effective potential.
 The detailed calculations are presented
 in Appendix C. The final result reads 
 
 \begin{equation}
  \Gamma(M,\mathcal{E},T,\mu)=\frac{N_c}{4\pi}\sum_{f}\mathcal{E}_f^2\sum_{k=1}^{\infty}\frac{e^{-\frac{M_f^2\pi k}{\mathcal{E}_f} }}{(k\pi)^2},\label{decay}
 \end{equation}

\noindent where we need to perform the summation in the flavors $f=u,d,s$. As we will see, the entire dependence of external conditions in the Schwinger 
pair production enters just in the effective 
masses $M_f\equiv M_f(\mathcal{E},T,\mu)$.
%
%
%
%
%
%
\section{Regularization}
The term $\theta_f^{\mathcal{E}}$ of the effective potential involves an integral, eq.(\ref{thetaE}), which
is divergent and must be regularized. We use here the vacuum-subtraction scheme\cite{nosso1}

\begin{eqnarray}
  &&\overline{\theta}_f^{\mathcal{E}}=\theta_f^{\mathcal{E}}-\theta^{vac}_f-\theta^{field}_f,\label{thetaE4}\\
  &&\overline{\theta}_f^{\mathcal{E}}=-\frac{N_c}{8\pi^2}\int_0^{\infty}ds\frac{e^{-sM_f^2}}{s^3}
  \left[\mathcal{E}_fs\cot(\mathcal{E}_fs)-1+\frac{(\mathcal{E}_f s)^2}{3}\right],\nonumber
\end{eqnarray}

\noindent where we have subtracted the vacuum contribution $\theta_f^{vac}$ given by

\begin{equation}
\theta_f^{vac}=\frac{N_c}{8\pi^2}\int_{0}^{\infty} ds \frac{e^{-sM_f^2}}{s^3},\label{3d}
\end{equation}
\noindent and a field contribution $\theta^{field}$ proportional to the energy
of the electric field $\sim eE^2$.  Since the NJL model in $3+1$ space-time dimensions
is not renormalizable, we should choose a regularization scheme. 
Here we adopt the $3$D-momentum cutoff to regularize eq.(\ref{3d})\cite{kleva,buballa} and we get
\begin{multline}
\theta^{vac}_f=-\frac{N_c}{8\pi^2}\left[M_f^4\ln\left(\frac{\Lambda+\sqrt{\Lambda^2+M_f^2}}{M_f}\right)-\right.\\
\left. \Lambda\sqrt{\Lambda^2+M_f^2}\left(M_f^2+2\Lambda^2\right) \right].\label{omegavac}
\end{multline}
For the condensates, we define the vacuum subtracted condensate as
\begin{eqnarray}
&&\overline{\phi}_f^{\mathcal{E}}= \phi_f^{\mathcal{E}}- \phi_f^{vac}    \label{condE4} \\ \nonumber
&&\overline{\phi}_f^{\mathcal{E}}=-\frac{MN_c}{4\pi^2}\int_0^{\infty}ds
\frac{e^{-sM^2}}{s^2}\left[\mathcal{E}_fs\cot(\mathcal{E}_fs)-1\right], 
\end{eqnarray}
\noindent where the vacuum contribution regularized with a 3D cutoff is given by

\begin{equation}
\phi_f^{vac}=-\frac{M_fN_c}{2\pi^2}\left[\Lambda \mathcal{E}_{\Lambda} -
      M_f^2\ln\left( \frac{\mathcal{E}_{\Lambda}+\Lambda}{M_f} \right)\right],
\end{equation}
where $\mathcal{E}_{\Lambda}=\sqrt{\Lambda^2+M^2_f}$.

The gap equation eq.(\ref{gapsu3}) in the NJL SU(3) should be regularized using the following regularized condensate $\phi_f^{\mathcal{E}}$
\begin{equation}
 \phi_f^{\mathcal{E}}=\overline{\phi}_f^{\mathcal{E}}+\phi_f^{vac},
\end{equation}
\noindent in the same way, for the effective potential eq.(\ref{omega1}), we should use the regularized $\theta_f^{\mathcal{E}}$
\begin{equation}
 \theta_f^{\mathcal{E}}=\overline{\theta}_f^{\mathcal{E}}+\theta_f^{vac}+\theta_f^{field}
\end{equation}

\noindent Although the integrals given in eqs.(\ref{thetaE4},\ref{condE4}) are already
regularized using the subtraction scheme in the vacuum, we still have poles associated to 
the zeros of $\sin(\mathcal{E}_fs)$ which appear in the denominator of
both our gap equation and the effective potential  
when $\mathcal{E}_fs=n\pi$ for $n=1,2,3,..$, 
and these poles will generate the imaginary part of the
effective potential that will be associated to the Schwinger pair production\cite{Schwinger}. 
For these reasons, these integrals should be
interpreted as Cauchy Principal Value\cite{ruggieri1}. Besides, in this work we are explicitly 
assuming that just the real values are present
in the gap equation. We assume this, since the real part of the effective potential is interpreted as the true ground state of the theory\cite{tatsumi}
and once the effective masses(and therefore the condensates) can be evaluated through the minimization of the effective potential, we can consider just their real values.

Using the analytical continuation technique, as discussed in the Appendix A, we can demonstrate 
that the Principal Value (or the real part) of 
$\overline{\theta}_f^{\mathcal{E}}$ is given by 
{\small 
\begin{eqnarray}
&&\Re \left( \overline{\theta}_f^{\mathcal{E}} \right) =-\frac{N_c}{2\pi^2}(\mathcal{E}_f)^2 
\left \{ \zeta'(-1) + \frac{\pi}{4}y_f  \right. \nonumber \\
&& + \frac{y_f^2}{2} ( \gamma_E-\frac{3}{2}+\ln y_f ) - \frac{1}{12} (1+ \ln y_f )     \label{potef1}  \\ 
&& \left. +\sum_{k=1}^{\infty}k\left[\frac{y_f}{k}\tan^{-1}\left(\frac{y_f}{k}\right)-
\frac{1}{2}\ln\left(1+\left(\frac{y_f}{k}\right)^2\right)-
\frac{1}{2}\left(\frac{y_f}{k}\right)^2\right] \right \}  ~,  \nonumber
\end{eqnarray} }
\noindent where $y_f=M_f^2/(2\mathcal{E}_f)$. Analogously we obtain 
for the Principal Value (or the real apart)
of the vacuum subtracted condensate $\overline{\phi}_f^{\mathcal{E}}$(see Appendix B)  
\begin{eqnarray}
\Re \left( \overline{\phi}_f^{\mathcal{E}} \right) =-\frac{M_fN_c}{4\pi^2}\int_0^{\infty}ds
\frac{e^{-sM_f^2}}{s^2}\left[\mathcal{E}_fs\cot(\mathcal{E}_fs)-1\right]   \nonumber \\ 
=\frac{M_fN_c}{2\pi^2}\mathcal{E}_f\left[\frac{\pi}{4}+y_{f}(\gamma_E-1+\ln y_{f})+\right. \nonumber \\
\left. +\sum_{k=1}^{\infty}\left(\tan^{-1}\frac{y_{f}}{k}-\frac{y_{f}}{k} \right)\right]. \label{reg1}
\end{eqnarray}

The quantities $\phi_f^{\mathcal{E},T,\mu}$ and $\theta_f^{\mathcal{E},T,\mu}$ depends on 
 temperature and chemical potential 
and following the authors of reference \cite{ayala}, we assume that the thermal part is
already regularized in the lower limit of the integration, i. e., we set the lower limits to zero,  
since theses integrals are finite. These temperature dependent quantities are evaluated 
through the numerical calculation of the integrals which appear in their explicit integral 
representations given in eq.(\ref{phi_tmu}) and eq.(\ref{thetaET}) and a rapid convergence is 
achieved with only a few terms summed. For the evaluation  of
$\phi_f^{\mathcal{E}}$ and $\theta_f^{\mathcal{E}}$ the corresponding analytical expressions are used.

%
%
%
%
%
%
\section{The two-flavor model}

In the NJL model with two flavors (SU(2) NJL) we have a lot of simplifications in our
previous equations. Let us start with the Lagrangian

\begin{multline}
\mathcal{L}=\overline{\psi}\left(i \slashed D - \tilde{m}\right)\psi
+G\left[(\overline{\psi}\psi)^{2}+(\overline{\psi}i\gamma_{5}\vec{\tau}\psi)^{2}\right]-
\frac{1}{4}F^{\mu\nu}F_{\mu\nu} ~,
\end{multline}

\noindent where, $\vec{\tau}$ are the isospin Pauli 
matrices, $Q$ is the diagonal quark charge 
 matrix, 
Q=diag($q_u$= $2 e/3$, $q_d$=-$e/3$), 
 $\psi=(u,d)^T$ is the quark fermion field, and $\tilde{m}=m_u=m_d$ represents the bare quark masses.

In the mean field approximation, the Lagrangian density reads
\begin{equation}
 \mathcal{L}=\overline{\psi}\left(i\slashed D-M\right)\psi+G \left \langle \overline{\psi}\psi \right \rangle^{2}-
 \frac{1}{4}F^{\mu\nu}F_{\mu\nu}~,
\end{equation}
\noindent where the constituent quark mass is defined by 

\begin{eqnarray}
M=m-2G\sum_{f=u,d} \phi_f, \label{gap} 
\end{eqnarray}

\noindent where we have used the definition given in eq.(\ref{cond1}). 

Now, using in the previous equation the regularized quantities given in 
the last section, the SU(2) NJL gap equation reads

\begin{equation}
 \frac{M-m}{2G}=-\sum_{f=u,d}\left(\overline{\phi}^{\mathcal{E}}_f+
 \phi_f^{\mathcal{E},T,\mu}+\phi_f^{vac}\right).\label{gap2}
\end{equation}

The Thermodynamical Potential is obtained just integrating eq. (\ref{gap2}) in the
effective mass $M$,

\begin{eqnarray}\label{pot1}
\Omega=\frac{(M-m)^2}{4G}-\sum_{f=u,d}\left(\overline{\theta}^{\mathcal{E}}_f+
\theta_f^{\mathcal{E},T,\mu} +\theta_f^{vac} \right),
\end{eqnarray}

\noindent where we are using all the definitions already presented in the section II. Notice that for the 
SU(2) NJL model $M_u=M_d=M$\cite{buballa}. 
This is a special situation that occurs only for the SU(2) NJL model
in the mean field approximation with equal current quark masses ($m_u=m_d$). In this case, the condensates 
contribute  symmetrically to the effective quark masses $M_u$ and $M_d$, i. e., 
$(M_f=m_f-2G\sum_{f=u,d} \phi_f)$. In a completely different way than in eq.(\ref{gapsu3}), 
the NJL SU(3) model will have three equations for each flavor. Hence,
we can expect an evident difference between the effective masses of the $u$ and $d$ quarks 
in the electric medium.
%
%
%
%
%
%
\section{Numerical Results}

 In the following we present the numerical results. For the SU(3) NJL model
we choose the following set 
of parameters: $\Lambda=631.4$MeV, $m_u=m_d=5.5$MeV,
$m_s=135$MeV, $G\Lambda^2=1.835$, $K\Lambda^5=9.29$ taken from \cite{buballa}. These parameters 
were fitted
to reproduce physical quantities as the pion decay constant $f_\pi=93.0$MeV, 
the pion mass $m_\pi=138$MeV and the chiral 
condensates $<\overline{u}u>^{\frac{1}{3}}=<\overline{d}d>^{\frac{1}{3}}=-246.9$ MeV,
$<\overline{s}s>^{\frac{1}{3}}=-267.0$MeV\cite{buballa}. In order to compare
more precisely and consistently the SU(3) and SU(2) NJL results, we have fitted the SU(2) NJL
model parameters to reproduce the same SU(3) physical values for $f_\pi$, $m_\pi$ and 
$<\overline{u}u>^{\frac{1}{3}}=<\overline{d}d>^{\frac{1}{3}}$ given above. 
Hence, our parameter set for the SU(2) NJL model are $\Lambda=632.66$MeV, $G\Lambda^2=2.17$ and
$m=5.38$MeV.

 We start showing the results for the effective quark masses as a function of the electric
field at fixed temperatures for both the SU(2) and SU(3)
versions of NJL model.

 From now on we will refer to the pseudocritical electric field, which is defined as the peak of minus the derivative of $M_i$ as a function of $eE$, i.e., $-\frac{dM_i}{d(eE)}$.
However, since in this work we are only interested in showing qualitatively the transition region, we do not evaluate such derivative.

\begin{figure}[!h]
\begin{tabular}{ccc}
\includegraphics[width=8.5cm]{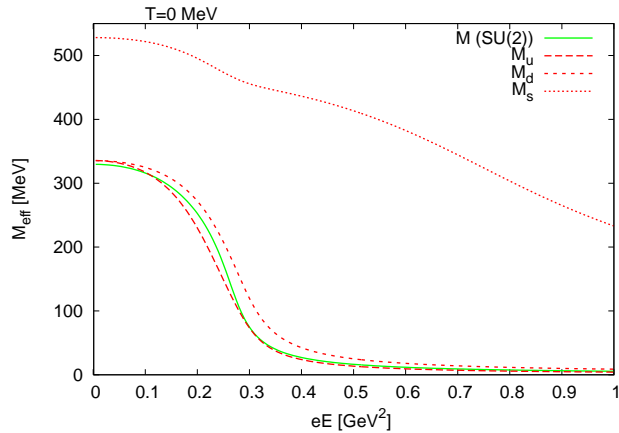}\\
\end{tabular}
\caption{Effective quark masses as a function of the electric field for $T=0$.}
\label{MET0}
\end{figure}

In Fig.\ref{MET0} we consider $T=0$ and one can observe 
the well-known behavior 
of the effective mass where the 
chiral symmetry is partially restored when a pseudocritical electric field $eE_c$ is reached. 
In the SU(3) version, the mass of the two lightest quarks $M_u$ and $M_d$  
 show a shift starting
at $eE\sim 0.1$GeV$^2$ due to the difference of the u and d quark electric charges. 
For electric fields larger 
than the critical
value, $E_c$, the $M_u$, $M_d$ and the SU(2) ,$M$, effective quark masses show  qualitatively the 
same behavior.
 The strange effective quark mass $M_s$ decreases much more slowly as a function of $eE$ when 
 compared to the masses of the quarks $u$, $d$ and $M$ and clearly the electric field necessary
 for the restoration of the chiral symmetry for the $s$ quark is much larger than the one 
 expected for the SU(2) version of the model.
The partial chiral symmetry restoration for the strange quark mass, i. e., when
$M_s\sim m_s$ occurs for a too strong electric field $eE>>\Lambda^2$ and 
we assume to be out a scope of  the NJL effective model.

\begin{figure}[!h]
\begin{tabular}{ccc}
\includegraphics[width=8.5cm]{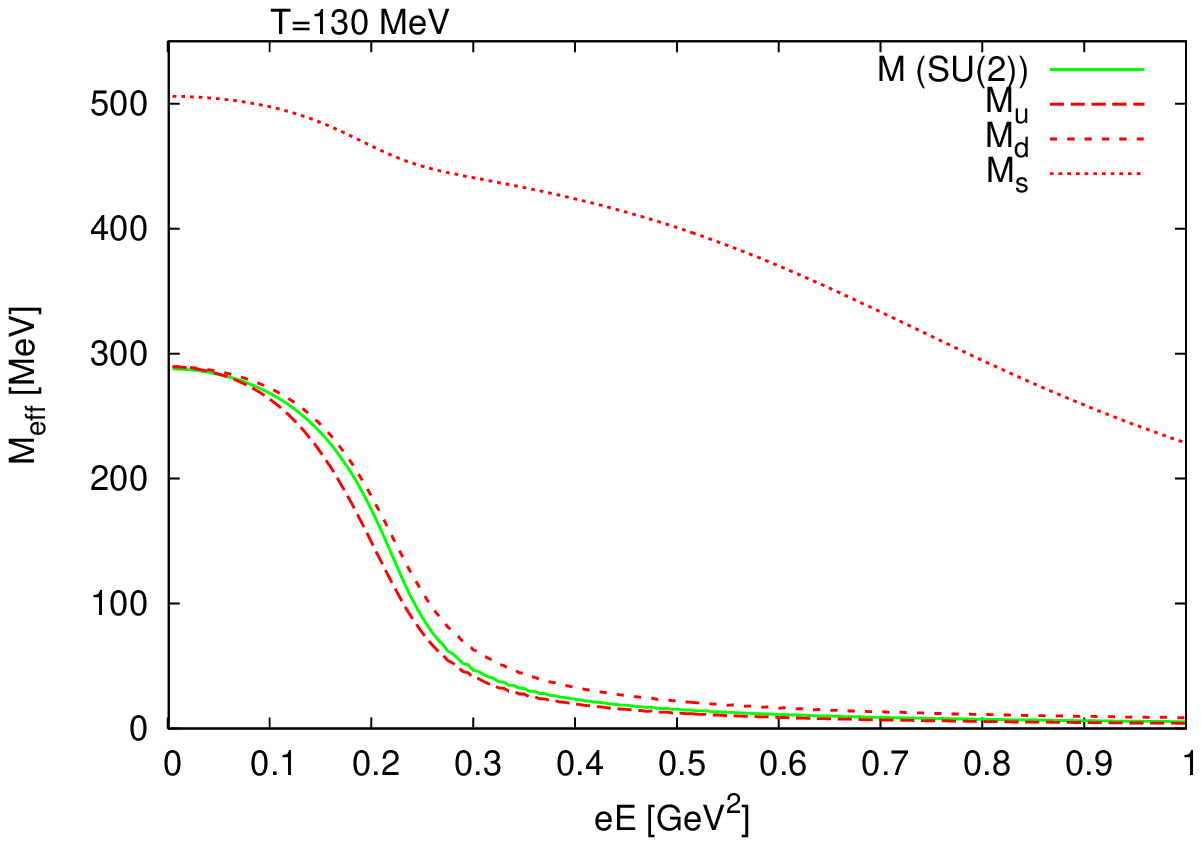}\\
\end{tabular}
\caption{Effective quark masses as a function of the electric field for $T=130MeV$.}
\label{MET130}
\end{figure}

\begin{figure}[!h]
\begin{tabular}{ccc}
\includegraphics[width=8.5cm]{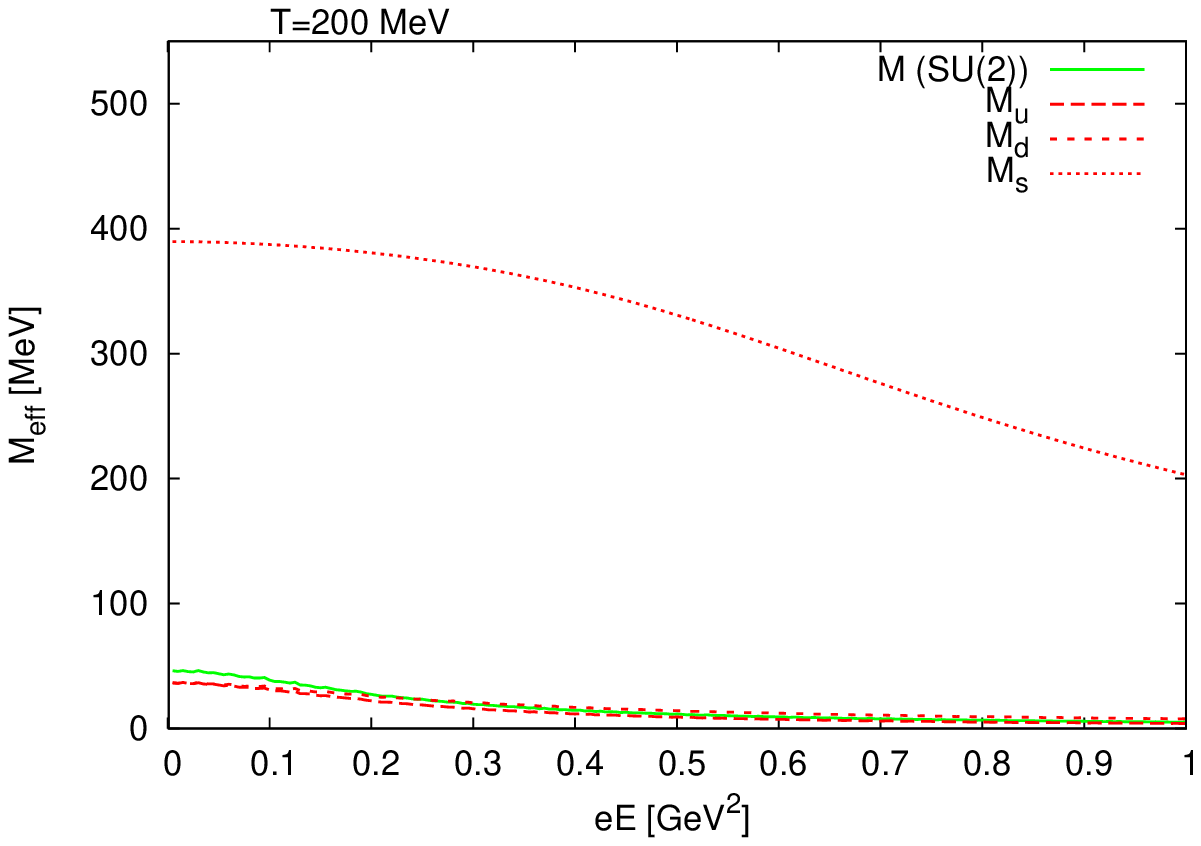}\\
\end{tabular}
\caption{Effective quark masses as a function of the electric field for $T=200MeV$.}
\label{MET200}
\end{figure}

 We next consider the effect of the temperature on the effective quark masses.
 In Fig.\ref{MET130} the effective masses are plotted as a function of $eE$ at $T=130$MeV.
 One can see that the temperature has the effect to break the chiral condensates
and just like electric field to weaken the constituent dynamical quark masses. 
In this way, we can see that when the  temperature grows 
the pseudocritical electric field $eE_c$ decreases. 
The same analysis can be done in Fig.\ref{MET200} where we fix $T=200$MeV.
 Here we notice that just due to the effect of the temperature  the chiral symmetry is almost 
completely restored and we can see a noticeable decrease of the pseudocritical electric field
by the order $eE_c\sim 0.15$GeV$^2$. 

\begin{figure}[!h]
\begin{tabular}{ccc}
\includegraphics[width=8.5cm]{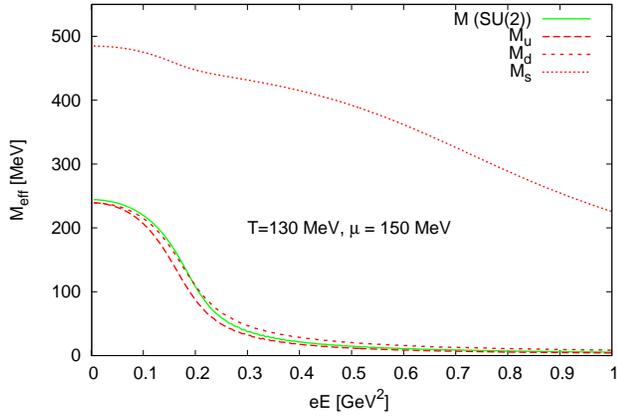}\\
\end{tabular}
\caption{Effective quark masses as a function of the Electric Field for $T=130$MeV and $\mu=150$MeV.}
\label{MTE130u75}
\end{figure}

In Fig.\ref{MTE130u75} we show the effect of finite chemical potential in the electrized quark matter 
for the SU(2) and SU(3) NJL models. In general,
 the chemical potential has the effect of partially restore the chiral symmetry, 
 weakening the effective masses in both models at $eE=0$. Hence, 
 we can expect the lowering of the critical electric field when the chemical 
 potential increases. 
 The effective masses of the lightest quarks have a similar 
 behavior in both models, with a natural displacement in the SU(3) effective masses
 at $eE>0.1$GeV$^2$ 
 due to the  difference of the $u$ and $d$ quark electric charges 
 and the strange quark effective mass is weakened as an effect of finite $\mu$. 

\begin{figure}[!h]
\begin{tabular}{ccc}
\includegraphics[width=8.5cm]{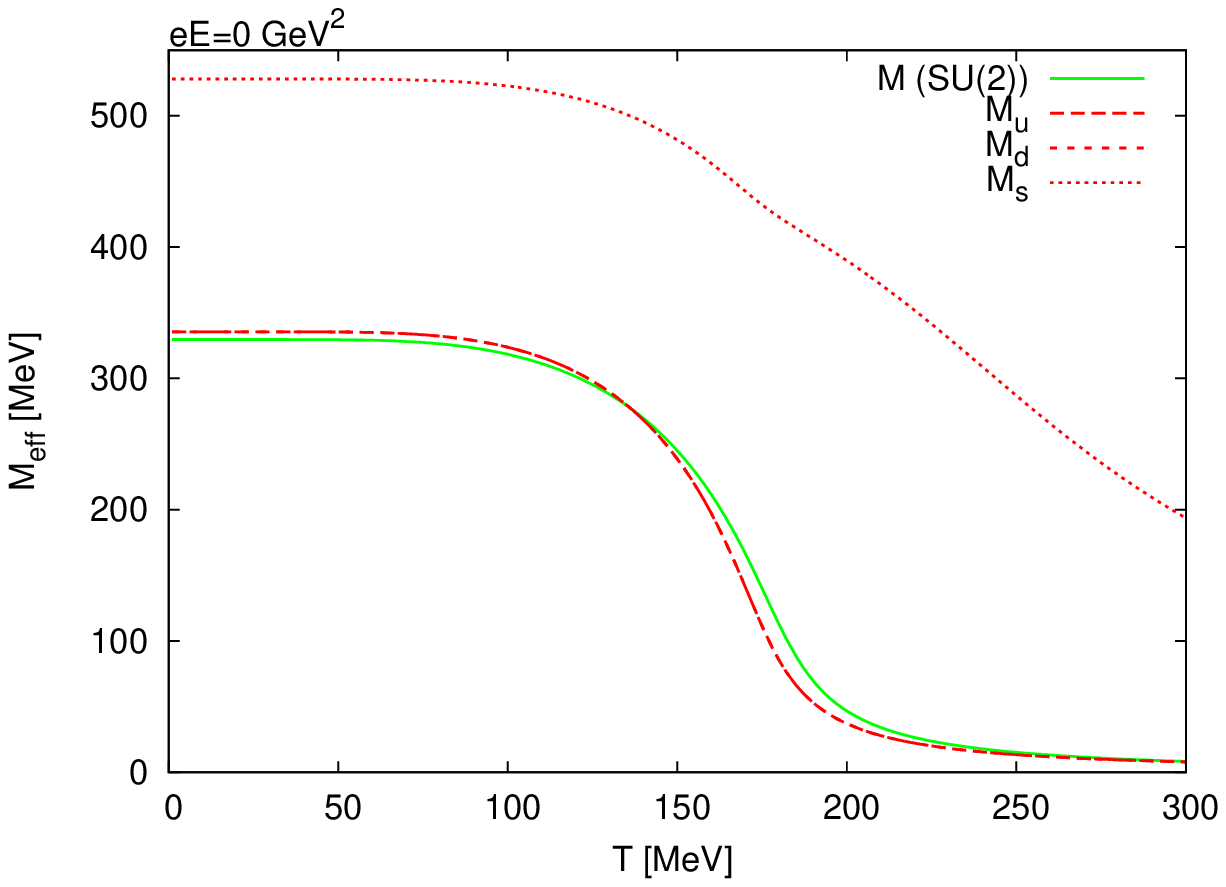}\\
\end{tabular}
\caption{Effective quark masses as a function of the temperature for the electric field $eE=0$.}
\label{MTE0}
\end{figure}

 In Fig.\ref{MTE0}, where the effective quark masses at $eE=0$ are plotted as a function of 
the temperature , the chiral symmetry restoration at finite temperature and zero electric field
can be analyzed. One can see that the restoration occurs around $T_c\sim 200$MeV and 
the behavior of the SU(2) and SU(3) light masses are qualitatively the same.

\begin{figure}[!h]
\begin{tabular}{ccc}
\includegraphics[width=8.5cm]{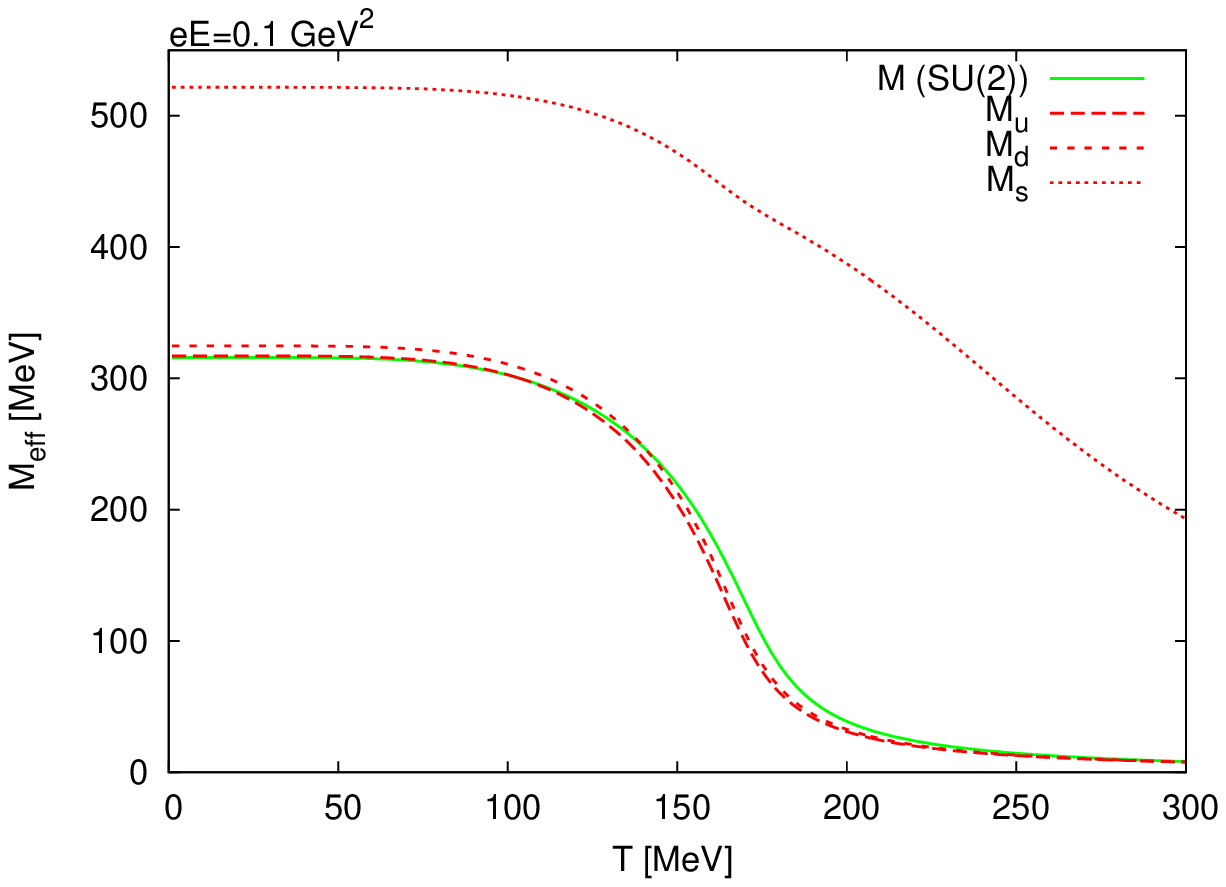}\\
\end{tabular}
\caption{Effective quark masses as a function of the temperature for the electric field
$eE=0.1$GeV$^2$.}
\label{MTE01}
\end{figure}

\begin{figure}[!h]
\begin{tabular}{ccc}
\includegraphics[width=8.5cm]{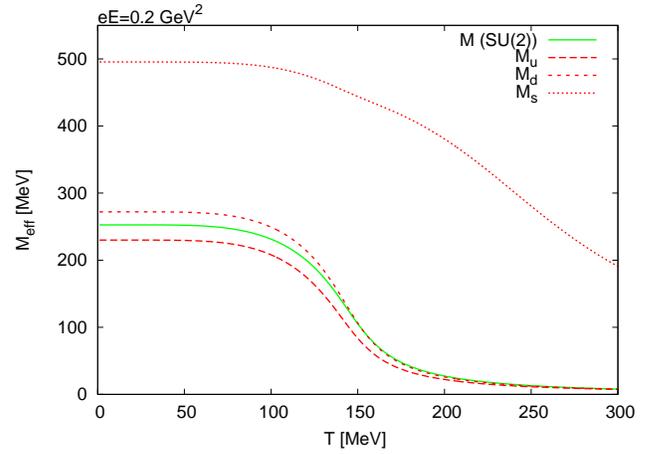}\\
\end{tabular}
\caption{Effective quark masses as a function of the temperature for the electric field $eE=0.2$GeV$^2$.}
\label{MTE02}
\end{figure}

The strange quark mass decreases 
more slowly, presenting a smooth bump at $T\sim 170$ MeV. In Fig.\ref{MTE01} one can see that 
the effect of the inclusion of a electric 
field $eE=0.1$GeV$^2$ is to decrease the effective mass of the strange quark and
 cause a shift of the effective $u$ and $d$ quark masses with the $u$ quark mass becoming larger than 
the $d$ quark mass. One can observe that both the electric field and the temperature 
weaken the quark condensates, however,
at sufficiently high temperatures the behavior of the lightest quark masses is 
qualitatively the same.
 In  Fig.\ref{MTE02}, as an effect of a stronger electric field, one can see a larger 
 shift of the lightest
effective quark masses and a slightly smaller effective strange quark mass.

From the figures \ref{MTE0},\ref{MTE01},\ref{MTE02} it is interesting to see that 
the (pseudocritical) temperature of the second order phase transition 
decreases with the increase
of the electric field, so the electric field enhances the chiral symmetry restoration. 
As mentioned earlier, if we increase the electric field, 
the imaginary part of the effective potential becomes different of zero and we can
associate this imaginary component to the creation of quark-antiquark pairs.

\begin{figure}[!h]
\begin{tabular}{ccc}
\includegraphics[width=8.5cm]{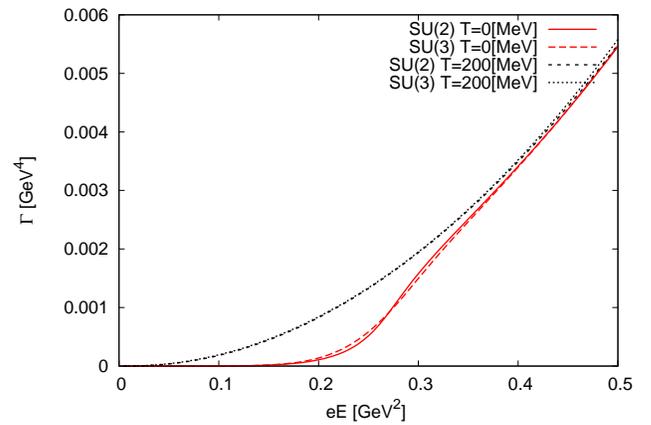}\\
\end{tabular}
\caption{Schwinger Pair-Production as a function of the electric field for temperatures
T=0 MeV and T=200 MeV.}
\label{GammaE}
\end{figure}

\begin{figure}[!h]
\begin{tabular}{ccc}
\includegraphics[width=8.5cm]{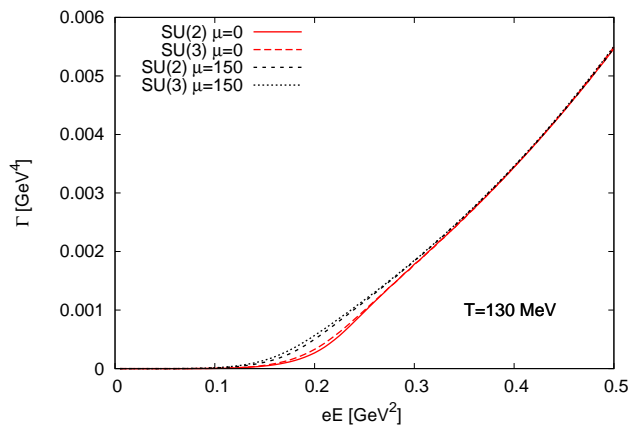}\\
\end{tabular}
\caption{Schwinger Pair-Production as a function of the electric field at $T=130$ MeV and
$\mu=0$ MeV and $\mu=150$ MeV.}
\label{GammaEmu}
\end{figure}

The Schwinger pair-production rate $\Gamma$ is shown 
in Fig.\ref{GammaE} as a function of the electric field for the 
two versions of the NJL model  at $T=0$ and $T=200$MeV.
The results shows very little difference 
between both models at $T=0$ and after $eE\sim0.2$GeV$^2$ the production rate 
grows more quickly due to the weakening of the chiral condensates and  the QCD vacuum
becomes more and more unstable and the pair of particle-antiparticle becomes more
likely to happen. 
If we rises the temperature to $T=200$MeV, we can see almost
no difference between the two models and
the production rate increases considerably for electric fields $eE<0.3$GeV$^2$ when compared to the case $T=0$.
The effect of finite chemical potential is shown in Fig.\ref{GammaEmu}, where we compare 
the production rate in both models with $\mu=0$ and $\mu=150$MeV at
$T=130$MeV. The two versions of the NJL model agree in their general aspects,
with quantitative differences in the transition 
region. As we can see
the effect of finite chemical potential is to increase slightly the production rate at 
lower electric field.

\begin{figure}[!h]
\begin{tabular}{ccc}
\includegraphics[width=8.5cm]{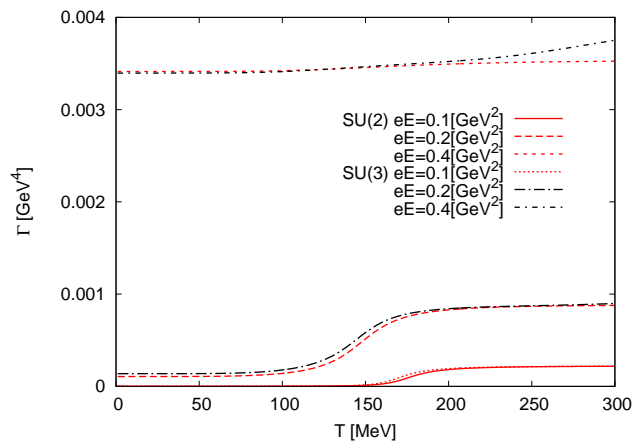}\\
\end{tabular}
\caption{Schwinger Pair-Production rate as a function of the temperatures for the electric 
fields $eE=0.1$GeV$^2$, $eE=0.2$GeV$^2$ and $eE=0.4$GeV$^2$.}
\label{GammaT}
\end{figure}

 In Fig.\ref{GammaT} we show the Schwinger pair production as a function of the 
temperature at fixed electric fields $eE=0.1$GeV$^2$, $eE=0.2$GeV$^2$ and $eE=0.4$GeV$^2$. At $eE=0.1$GeV$^2$ we can see that
the production rate grows quickly when a phase transition becomes more apparent at $T\sim 150$MeV, 
with a more prominent production rate for the SU(3) model in comparison to the SU(2)
and stabilizes at $T\sim 200$MeV. This happens because the phase transition in this case is driven entirely by
the temperature and when the chiral symmetry is partially restored we can expect the Schwinger pair production to
become  almost stable. If we increase the
electric field to $eE=0.2$GeV$^2$, the production rate is more significant in the SU(3) model
and when we reach $T\sim100$MeV the production rate starts to increase
more quickly and stabilize again, when for the two NJL models the Schwinger rate almost coincides.
However, the production rate is more than four times greater than the production rate of 
$eE=0.1$GeV$^2$ case. 

We also show our results for $eE=0.4$ GeV$^2$ since this value is approximately 
the electric field predicted in the simulations\cite{deng}. For this electric field
the chiral symmetry has already been partially restored
and almost no quantitative difference is seen for $T<200$MeV in both models, but the
effects of high temperatures
are prominent in the SU(3) model where the production rate grows 
while in the SU(2) the production rate stabilizes. 

\section{Conclusions}

In this work we use the SU(2) and SU(3) versions of Nambu--Jona-Lasinio model at finite temperature and densities
to study how a constant electric field in the $z$ direction
can affect the chiral symmetry restoration. To this end, in the the SU(3) version
we improve the calculations by including the 
't Hooft determinant in comparison with\cite{tatsumi} and also assuming, 
differently from ref.(\cite{Cao}), non-zero current quark masses 
in both SU(2) and SU(3) models in order to calculate the
effective quark masses and the Schwinger pair production. 

 The real part of the gap equation and of the effective potential 
should be properly regularized,  since their $T=0$ contributions are divergent.
We derive a set of regularized expressions 
obtained by analytical continuation in the Appendices of this work. These expressions  
are much more convenient to be used in numerical calculations, since avoids
the highly-oscillatory integrals of eq.(\ref{phisu3}) and 
eq.(\ref{thetaE})\cite{Cao,ruggieri1,ruggieri2,tatsumi} and as usual for the expressions
at finite $T$ and $\mu$ we do not use any regularization
since these integrals are finite\cite{ayala}.

Firstly, we explore how the electric field restores the chiral symmetry. The general
feature of the electric field is to the break the chiral condensates and in comparison
to the SU(2) case, we can see a splitting of the dynamically generated  
masses of the $u$ and $d$ quarks $M_u$ and $M_d$ at relatively weak electric fields $eE\sim 0.1$GeV$^2$. 
For the strange quark, its effective mass $M_s$ decreases more slowly and the current
quark mass $m_s$ is reached only at a very strong electric fields. 
 The net effect is that the higher is the electric field the lower is the (pseudocritical)
temperature of chiral restoration.

Analogously, the effect of the temperature is to enhance the chiral symmetry restoration 
and the higher the temperature the lower the corresponding electric field where the 
chiral symmetry is restored.
The results for the Schwinger pair production evaluated in the SU(2) and SU(3) versions of the NJL model 
show similar behavior, at low temperatures and 
with the electric field rising,
the production tends to increase when we cross a pseudocritical electric field and
if we increase the temperature, the Schwinger pair production tends to
initiates at lower electric fields. Besides, the inclusion of strange quark matter
in this work indicates that the production of quark-antiquark pairs should be 
more pronounced when compared with the results of the SU(2) model.

%
%
%
%
%
\appendix
\section{ The Principal Value of $\overline{\theta}^{\mathcal{E}}_f$}

To evaluate the Principal Value of $\overline{\theta}^{\mathcal{E}}_f$ as given in eq.(\ref{thetaE4}), we will 
perform an analytical continuation from the closely related expression
obtained in the framework of magnetized quark matter subjected to a constant magnetic
field in the $z$ direction in the 
context of NJL model. 
In this way, we just start writing the well-known result\cite{avancini2,andersen,dunne}

\begin{eqnarray}
&&\overline{\theta}^{\beta}_f =  -\frac{N_c}{8\pi^2} \int_0^{\infty}ds \frac{e^{-sM_f^2}}{s^3}
\left[\beta_f s \coth(\beta_f s)-1 - \frac{\beta_f^2}{3}  \right]  \nonumber \\
  &=& \frac{N_c\beta_f^2}{2\pi^2}\left[\zeta'(-1,x_f)-\frac{1}{2}(x_f^2-x_f)\ln x_f+
\right.  \nonumber\\
&& \left. \frac{x_f^2}{4}-\frac{1}{12}(1+\ln x_f )\right]\label{ap1}  ~,
\end{eqnarray}

\noindent where $x_f=\frac{M_f^2}{2\beta_f}$ and $\beta_f=|q_f|B$. The duality between magnetic 
and electric fields can 
be seen through the replacement $eB\rightarrow -ieE$\cite{Cao,kleva}. Therefore, we perform this 
duality in eq.(\ref{ap1}) through the
 prescription $x_f\rightarrow iy_f$ 
\begin{eqnarray}
\overline{\theta}^{\mathcal{E}}_f= -\frac{N_c\mathcal{E}_f^2}{2\pi^2}\left[\zeta'(-1,iy_f)-
\frac{1}{2}(-y_f^2-iy_f)\ln(iy_f)\right.\label{theta5}\nonumber\\
\left. -\frac{y_f^2}{4}-\frac{1}{12}(1+\ln(iy_f)  )\right]\label{potef2} ~.
\end{eqnarray}

Now the main difficulty is to evaluate $\zeta'(-1,iy_f)$, since the remaining terms are
almost trivial to obtain. 
To proceed we use a convenient relation between the derivative of the Riemann zeta function and the 
logarithm of the gamma function, $\ln \Gamma$,  quoted without proof in ref.\cite{dunne}.
\begin{equation}
\zeta'(-1,x)=\zeta'(-1)-\frac{x}{2}(1-x)-\frac{x}{2}\ln2\pi+\int_0^xdx'\ln\Gamma(x') \label{zetaprime} ~.
\end{equation}
Next we sketch a proof of the latter expression. Firstly, we write 
\begin{equation}
 \frac{\partial\zeta'(-1,x)}{\partial x} \equiv
 \left. \frac{\partial}{\partial x}\frac{\partial}{\partial z} \zeta(z,x)\right|_{z=-1} =
 \left. - \frac{\partial}{\partial z} z \zeta(z+1,x)\right|_{z=-1} \label{zetder} ~ ,
\end{equation}
where the last equality follows from the relation\cite{apostol}
\begin{equation}
 \frac{\partial}{\partial x} \zeta(z,x) = - z\zeta(z+1,x) \nonumber ~.
\end{equation}
We now calculate the last derivative in eq.(\ref{zetder}) and use the following equalities\cite{apostol}
\begin{equation}
\zeta(0,x) = \frac{1}{2} -x~~,~~  \zeta'(0,x)= \ln \Gamma (x) - \frac{1}{2} \ln(2\pi) \nonumber ~ ,
\end{equation}
thus obtaining the expression:
\begin{equation}
 \frac{\partial\zeta'(-1,x)}{\partial x}=\left(\frac{1}{2}-x\right)+
 \left[\ln\Gamma(x)-\frac{1}{2}\ln(2\pi)\right]\nonumber ~.
\end{equation}
\noindent A simple integration of the latter equation yields 
\begin{equation}
\zeta'(-1,x)=\zeta'(-1)-\frac{x}{2}(1-x)-\frac{x}{2}\ln2\pi+\int_0^xdx'\ln\Gamma(x')\label{zetprime2} ~,
\end{equation}

\noindent with $\zeta'(-1)=\frac{1}{12}-\ln(A)$, where $A=1.2814271291...$, the Glaisher–Kinkelin 
constant\cite{glaisher}. To evaluate the integral
that has been left in the last equation we need to invoke a representation of $\ln \Gamma(x) $\cite{boros}

\begin{equation}
\ln\Gamma(x)=-\gamma_Ex-\ln(x)+\sum_{k=1}^{\infty}\left[\frac{x}{k}-\ln(1+\frac{x}{k})\right]\label{logama},
\end{equation}

\noindent where $\gamma_E$ is the Euler-Mascheroni constant. 
Integrating the latter equation over the variable $x$ from $0$ to $i y_f$, one obtains 
the analytical continued expression
{\small 
\begin{eqnarray}
 &&\int_0^{iy_f}dx\ln\Gamma(x)=\frac{\gamma_Ey_f^2}{2}+y_f\frac{\pi}{2}+\sum_{k=1}^\infty\left[-\frac{y_f^2}{2k}-
 \right.\nonumber\\
 &&\left. k\left(\frac{1}{2}\ln\left(1+\frac{y_f^2}{k^2}\right)-\frac{y_f}{k}\tan^{-1}\frac{y_f}{k}\right)\right]
 +\nonumber\\
 && i\left \{-y_f\ln y_f+y_f \right. \nonumber\\
 &&\left. -\sum_{k=1}^{\infty}\left[k\left(\frac{y_f}{2k}\ln\left(1+\frac{y_f^2}{k^2}\right)+
 \tan^{-1}\frac{y_f}{k}\right)-y_f \right] \right \}\label{lng}.
\end{eqnarray}  }
From eqs.(\ref{potef2},\ref{zetprime2},\ref{lng}) the analytically continued 
expression as given in eq.(\ref{potef1})  can be straightforwardly obtained. 

%
%
%
%
\section{ The Principal Value of $\overline{\phi}_f^{\mathcal{E}}$}
Next, we apply the same analytical continuation technique
of the previous appendix in order to derive the principal value of the quark condensates. We use
the SU(3) NJL condensates in a magnetic field taken from the literature\cite{nosso1,avancini2}.
To this end, we start from the regularized part of the condensate in a magnetic field  

\begin{eqnarray}\label{mag1}
\overline{\phi}^{\beta}_f=-\frac{N_c M_f}{4\pi^2}\beta_f\int_0^{\infty}ds\frac{e^{-sM_f^2}}{s}
\left[\coth(\beta_fs)-1\right]  \nonumber \\
=-\frac{N_cM_f}{2\pi^2} \beta_f\left[\ln\Gamma(x_{f})-\frac{1}{2}\ln(2\pi)\right.  \nonumber \\
\left. +x_f-\frac{1}{2}\left(x_{f}-1\right)\ln(x_{f})\right].
\end{eqnarray}

\noindent So, performing the analytic continuation of the latter equation,
we can write the dual expression for the condensate, $\overline{\phi}^{\mathcal{E}}_f$, in a constant electric
field in the $z$ direction as

\begin{eqnarray}\label{elec1}
\overline{\phi}^{\mathcal{E}}_f=\frac{-iN_cM_f}{4\pi^2}\mathcal{E}_f\int_0^{\infty}ds\frac{e^{-sM_f^2}}{s}
\left[\coth(-i\mathcal{E}_fs)-1\right]   \nonumber \\
=-\frac{M_fN_c}{2\pi^2} i\mathcal{E}_f\left[\ln\Gamma(iy_{f})-\frac{1}{2}\ln(2\pi)\right.  \nonumber \\
\left. +iy_f-\frac{1}{2}\left(iy_{f}-1\right)\ln(iy_{f})\right]. 
\end{eqnarray}

We are interested in the real part of the eq.(\ref{elec1}). Since $\coth(-ix)=i\cot(x)$, we have

\begin{eqnarray}
\Re (\overline{\phi}^{\beta}_f)=\Re\left\{\frac{N_cM_f}{4\pi^2}\mathcal{E}_f\int_0^{\infty}ds\frac{e^{-sM_f^2}}{s}\left[\cot(\mathcal{E}_fs)-1\right] \right\}\nonumber\\ 
=-\frac{N_cM_f}{2\pi^2} \mathcal{E}_f\Re \left[i\ln\Gamma(iy_{f})-i\frac{1}{2}\ln(2\pi)+\right.\nonumber \\ 
\left. -y_f-i\frac{1}{2}\left(iy_{f}-1\right)\ln(iy_{f})\right] ~.  \label{elec2}
\end{eqnarray}

The imaginary part of $\ln\Gamma(x)$ can be extracted from its series representation 
given by eq.(\ref{logama}). If $z=|z| e^{i\theta}$ is an arbitrary complex number, its logarithm is given by
$\ln z = \ln|z| + i \theta$. Using this expression one easily obtains:
\begin{equation}
 \ln(1+i\frac{x}{k})=\frac{1}{2} \ln \left(1+\frac{x^2}{k^2} \right) + i\tan^{-1}\frac{x}{k} ~,
\end{equation}
from the latter equation it follows that:
\begin{eqnarray}
 \Im \left[\ln(1+i \frac{y_f}{k})\right]=\tan^{-1}\frac{y_f}{k} \nonumber ~.
\end{eqnarray}
After substituting the last equation in eq.(\ref{logama}), it is straightforward to show that:
\begin{equation}
\Im \left( \ln\Gamma(iy_f) \right)=
-\gamma_E y_f-\frac{\pi}{2}+\sum_{k=1}^{\infty}\left[\frac{y_f}{k}-\tan^{-1}(\frac{x}{k})\right]~,
\end{equation}
The imaginary part of the eq.(\ref{elec1}) can now easily be obtained, hence, the desired eq.(\ref{reg1}) 
can be derived.
\section{Derivation of the Schwinger pair production rate $\Gamma$}
\noindent We have to extract the imaginary part of the effective potential in order to obtain  
the Schwinger pair production rate, eq.(\ref{decay}).Since we wish to evaluate 
the imaginary part of $\theta^{\mathcal{E}}_f$,
the term proportional to $\mathcal{E}^2$ can be discarded, and the eq.(\ref{thetaE}) should be given by
\begin{equation}
\overline{\theta}_f^{\mathcal{E}}=-\frac{N_c}{8\pi^2}\int_0^{\infty}ds\frac{e^{-sM_f^2}}{s^3}
\left[\mathcal{E}_fs\cot(\mathcal{E}_fs)-1\right].\label{thetaE2}
\end{equation}
We now use the following trigonometric relation \cite{remment}
\begin{equation}
 \pi\cot(\pi x)=\frac{1}{x}+\sum_{k=1}^\infty\frac{2x}{x^2-k^2},
\end{equation}
\noindent in eq.(\ref{thetaE2}).
After an appropriate change of variable, one obtains
\begin{equation}
\overline{\theta}_f^{\mathcal{E}}=-\frac{N_c}{8\pi^4}\mathcal{E}_f^2\int_0^{\infty}ds\frac{e^{-s\frac{\pi M_f^2}{\mathcal{E}_f}}}{s^3}\left(\sum_{k=1}^{\infty}\frac{2s^2}{s^2-k^2}\right).\label{thetaE3}
\end{equation}

\noindent  Next, performing a simple partial fraction decomposition and using the identity 
$\lim_{\epsilon \rightarrow 0} \frac{1}{x\pm i\epsilon}=P.V\frac{1}{x}\mp i\pi\delta(x)$, we obtain

\begin{equation}
\Im({\overline{\theta}_f^{\mathcal{E}}})=\frac{N_c}{8\pi}\mathcal{E}_f^2\sum_{k=1}^{\infty}\frac{e^{-k\frac{\pi M_f^2}{\mathcal{E}_f}}}{(\pi k)^2},\label{decay2}
\end{equation}

\noindent plugging this result in the usual definition of the Schwinger pair-production rate\cite{Schwinger,Cao} 
, $\Gamma=-2\Im{(\Omega)}$, we obtain eq.(\ref{decay}).

\section{Equivalence between the imaginary part of eq.(\ref{potef2}) and eq.(\ref{decay2}) }

Making use of eq.(\ref{potef2},\ref{zetprime2}) and a change of variables $x'=ix$ in the integration, the full imaginary part of the effective potential is given by 

\begin{eqnarray}
\Im{(\overline{\theta}^{\mathcal{E}}_f)}&=& -\frac{N_c\mathcal{E}_f^2}{2\pi^2}\left[\Re\int_0^{y_f}dx\ln\Gamma(ix)
-\frac{y_f}{2} -\frac{y_f}{2}\ln(2\pi) \right. \nonumber\\&+&\frac{y_f}{2}\ln y_f
\left. -\frac{y_f}{2} -\frac{y_f}{2}\ln(2\pi)+\frac{\pi}{4}y_f^2-\frac{\pi}{24} \right].\label{log2}
\end{eqnarray}

Now, to proceed we have to integrate $\ln\Gamma(x)$. First, we make use of the property $x\Gamma(x)=\Gamma(x+1)$, to obtain
{\small
\begin{eqnarray}
 \Re\int_0^{y_f}dx\ln\Gamma(ix)=\Re\left[\int_0^{y_f}dx\ln\Gamma(1+ix)-\int_0^{y_f}dx\ln(ix)\right]\label{log1},
\end{eqnarray}}

\noindent the second integration is trivial, and is given by

\begin{equation}
\int_0^{y_f}dx\ln(ix)=y_f(\ln y_f-1)+\frac{i\pi}{2}y_f.
\end{equation}

By using the formulas of references\cite{grad,dittrich} 

\begin{eqnarray}
\Re\left[\ln\Gamma(1+ix)\right]&&=\frac{1}{2}\ln|\Gamma(1-ix)\Gamma(1+ix)|\\
&&=-\frac{1}{2}\ln\frac{\sinh(\pi x)}{\pi x},\nonumber
\end{eqnarray}

\noindent and performing an integration by parts, the first integral in eq.(\ref{log1}) can be solved 

\begin{eqnarray}
 \Re\left[\int_0^{y_f}dx\ln\Gamma(1+ix)\right]=-\frac{1}{2}\left \{y_f\ln\frac{\sinh(\pi y_f)}{\pi y_f}\right.\\\nonumber
 \left.-\int_0^{y_f}dx\left[\pi x\coth(\pi x)-1\right] \right \},
\end{eqnarray}

\noindent performing of variables $\pi x=z$ and the following result\cite{dittrich}

\begin{eqnarray}
 \frac{1}{\pi}\int_0^{y_f\pi} dz z\coth(z) =\frac{1}{2\pi}\left[-(y_f\pi)^2+2y_f\pi\ln(2)\right.\\ \nonumber
 \left.+2y_f\pi\ln\sinh(\pi y_f)-Li_{2}(e^{-2y_f\pi})+Li_2(1)\right],
\end{eqnarray}

\noindent where $Li_{2}(x)=\sum_{k=1}^{\infty}\frac{x^k}{k^2}$ is the Polylogarithm function of order 2 and we also use the very famous result 
$Li_2(1)=\frac{\pi^2}{6}$. 
Using these results, all the remaining terms of eq.(\ref{log2}) cancel each other, remaining only

\begin{eqnarray}
\Im{(\overline{\theta}^{\mathcal{E}}_f)}= -\frac{N_c\mathcal{E}_f^2}{2\pi^2}\left[-\frac{1}{4\pi}Li_{2}(e^{-2y_f\pi}) \right]\\
=\frac{N_c\mathcal{E}_f^2}{8\pi}\sum_{k=1}^{\infty}\frac{e^{-k\frac{M_f^2\pi}{\mathcal{E}_f}}}{(\pi k)^2}.
\end{eqnarray}

\section*{Acknowledgments}

This work was partially supported by Conselho Nacional
de Desenvolvimento Cientifico e Tecnologico (CNPq)
Grant No. 6484/2016-1 and as a part of the project
INCT-FNA (Instituto Nacional de Ciência e Tecnologia -
Física Nuclear e Aplicações (INCT-FNA)) 464898/2014-5
(S. S. A) and Coordenacao de Aperfeicoamento de Pessoal
de Nivel Superior (CAPES) (WRT).

\end{document}